# The Fano and EIT-like resonance characteristic of asymmetric double micro-ring resonator

**Chaoying Zhao***

*College of Science, Hangzhou Dianzi University, Zhejiang, 310018 P. R. China*

**Corresponding author: zchy49@hdu.edu.cn*

By breaking the symmetrical arrangement of double micro-ring resonator, the formation mechanism and performance of reflection and transmission spectrum and optical field distribution are investigated. The reflection spectrum is Fano shape. The Fano resonance has an asymmetric and sharp resonance peak can be independently tuned by changing the asymmetric coupling factor of the first micro-ring. The transmission spectrum is electromagnetically induced transparency (EIT)-like shape. The (EIT)-like resonance can be independently tuned by changing the absorption factor and the phase shift factor of the second micro-ring. The Fano and EIT-like resonance have low loss and high near-field localization characteristics, our research has promising applications in optical communication and optoelectronic modulators.

## 1. INTRODUCTION

As we all know, the coupled-mode theory can be used in the study of the performance of single and double- micro ring resonators (MRRs)[1-7]. We noticed that the interaction between double MRR is ignored [5-6]. Meng *et. al.* firstly studied the transmission spectra of $N \times N$ weak linearly array when the interaction between micro-ring (MR) is to be considered [8]. Considering many atomic coherent effects such as electromagnetic induction transparency (EIT) can be realized through all-optical method. Zheng *et.al.* firstly observed light EIT-like phenomenon in a controlled double MR coupling system [7]. Xiao *et.al.* realized the tunneling induced transparency effect in chaotic optical micro-cavity[9]. We find that compact silicon-on-insulator asymmetric embedded dual micro-ring resonators can produce the Fano spectrum[10]. In 2021, Zhao *et.al.* have studied dual-band electromagnetically induced transparency (EIT)-like effect in terahertz spectrum based on the traveling waves coupling mechanism[11]. In 2022, Zhao proposed a new counter-propagation waves coupling mechanism[12], and find out that the transparency window breadths of transmission spectra are greatly enhanced and the corresponding phase shift spectra possess a flat profile or a square profile. How to convert between EIT-like spectrum and Fano spectrum? A more generally study of the interaction between the waveguide and multi-MR system is needed.

## 2. ASYMMETRIC DOUBLE MICRO-RING RESONATOR

As shown in Fig.1(a), our scheme guarantee that $a_1 \to b_1, a_2 \to b_2 \to b_3 \to a_3 \to a_2$ and $a_4 \to b_4$ in the same direction from left to right, one bus waveguide $a_1 \to b_1$ and MR1 $a_2 \to b_2 \to b_3 \to a_3 \to a_2$, the MR2 $a_4 \to b_4$. As shown in Fig.1(b), when light field inject into bus wave-guide, the light field propagates through MR1 and MR2 in a clockwise direction. Therefore, the interaction in MRs produces the electromagnetically induced transparency (EIT) -like effect.

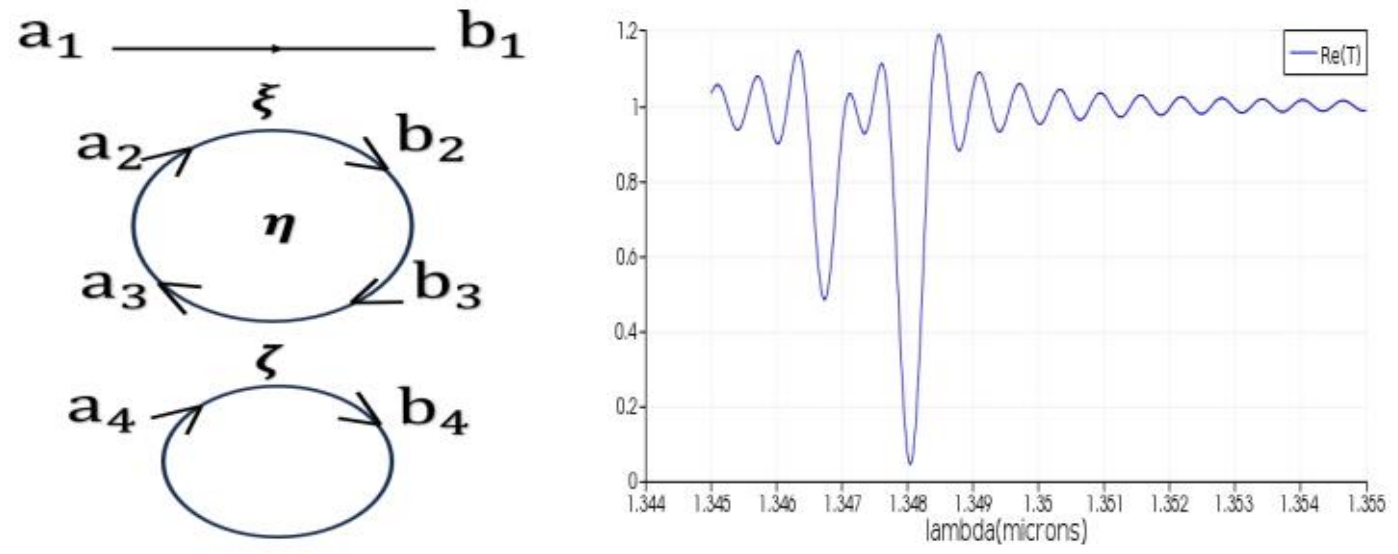

**Fig. 1.** (a)The asymmetric coupler, (b)The transmission spectrum.

The asymmetric coupling factor $\xi$ between the bus waveguide and the upper halves of MR1, the asymmetric coupling factor $\eta$ between the upper and the lower halves of the MR1, the asymmetric coupling factor $\varsigma$ between the lower halves of the MR1 and the upper of the MR2. The asymmetric weekly linearly coupling mode equations can be described by[ 8,10-12,14]

$$\begin{pmatrix} \frac{\partial a_1}{\partial z} \\ \frac{\partial a_2}{\partial z} \\ \frac{\partial a_3}{\partial z} \\ \frac{\partial a_4}{\partial z} \end{pmatrix} = \begin{pmatrix} 0 & j\frac{K}{2}\xi & 0 & 0 \\ j\frac{K}{2}\xi & 0 & j\frac{K}{2}\eta & 0 \\ 0 & j\frac{K}{2}\eta & 0 & j\frac{K}{2}\varsigma \\ 0 & 0 & j\frac{K}{2}\varsigma & 0 \end{pmatrix} \begin{pmatrix} a_1 \\ a_2 \\ a_3 \\ a_4 \end{pmatrix}, \qquad (1)$$

where $K/2$ denotes the symmetric coupling coefficient between adjacent waveguides and $a_i$ is the mode fields in the i-th waveguides.

Setting the eigen-solution

$$\partial a_i / \partial z = -\lambda a_i \, , a_i = a_i(x) e^{-jxkz} = a_i(x) e^{-jx\tau} \, , i = 1-4 \, ,$$

where $\tau = kz$ is the interaction strength between wave-guide and MRs,

Setting $a_i = a_{i0} e^{-\lambda z}$ in Eq.(1), the Eigenvalue-function can be derived by

$$Det = \begin{vmatrix} \lambda & j\frac{K}{2}\xi & 0 & 0 \\ j\frac{K}{2}\xi & \lambda & j\frac{K}{2}\eta & 0 \\ 0 & j\frac{K}{2}\eta & \lambda & j\frac{K}{2}\varsigma \\ 0 & 0 & j\frac{K}{2}\varsigma & \lambda \end{vmatrix} = \lambda^4 + \frac{K^2}{4}(\xi^2 + \eta^2 + \varsigma^2)\lambda^2 + \frac{K^4}{16}\xi^2\varsigma^2 = 0 \, . \qquad (2)$$

Setting $\lambda = jKx = j(K/2)2x$, the determinate of the Eigen-values of $2x$ satisfies

$$2x_1 = \sqrt{\frac{\xi^2 + \eta^2 + \varsigma^2}{2} + \sqrt{(\frac{\xi^2 + \eta^2 + \varsigma^2}{2})^2 - \xi^2\varsigma^2}} \, , x_2 = -x_1 \, ,$$

$$2x_3 = \sqrt{\frac{\xi^2 + \eta^2 + \varsigma^2}{2} - \sqrt{(\frac{\xi^2 + \eta^2 + \varsigma^2}{2})^2 - \xi^2\varsigma^2}} \, , \; x_4 = -x_3 \, ,$$

$$(2x_1)^2 (2x_2)^2 = \xi^2 \varsigma^2 \, . \qquad (3)$$

The eigenvectors can be written as $\varphi_i(x_m)$ (corresponding to the Eigen-value $x_m$ ), which is the components of the eigenvector

$$\varphi_i(x_m)=(\varphi_1(x_m),\varphi_2(x_m),\varphi_3(x_m),\varphi_4(x_m)),$$

the corresponding Eigen-functions

$$a_i(x_m)=\varphi_i(x_m)e^{ix_m\tau}.$$

If the eigenvalue $x_i$ are given, the corresponding eigenvector $\varphi_i(x_i)$, $i=1-4$ can be given out by the cofactor $D_{ij}$ [13-14](see Appendix A). Using the normalization

$$\sum_{i=1}^{4}\varphi_i^2(x_1)=1,\ \sum_{i=1}^{4}\varphi_i^2(x_4)=1,$$

and satisfy the following orthogonal relation

$$\varphi_1(x_1)\varphi_1(x_2)\varphi_1(x_3)\varphi_1(x_4)+\varphi_2(x_1)\varphi_2(x_2)\varphi_2(x_3)\varphi_2(x_4)$$
$$+\varphi_3(x_1)\varphi_3(x_2)\varphi_3(x_3)\varphi_3(x_4)+\varphi_4(x_1)\varphi_4(x_2)\varphi_4(x_3)\varphi_4(x_4)=0. \quad (4)$$

The determinate $\Delta$ can be expressed as the mix product of the vectors

$$\vec{a}(\varphi_1(x_1),\varphi_2(x_1),\varphi_3(x_1),\varphi_4(x_1)),\ \vec{b}(\varphi_1(x_2),\varphi_2(x_2),\varphi_3(x_2),\varphi_4(x_2)),$$
$$\vec{c}(\varphi_1(x_3),\varphi_2(x_3),\varphi_3(x_3),\varphi_4(x_3)),\ \vec{d}(\varphi_1(x_4),\varphi_2(x_4),\varphi_3(x_4),\varphi_4(x_4)).$$

$$\Delta=\begin{vmatrix}\varphi_1(x_1) & \varphi_1(x_2) & \varphi_1(x_3) & \varphi_1(x_4)\\ \varphi_2(x_1) & \varphi_2(x_2) & \varphi_2(x_3) & \varphi_2(x_4)\\ \varphi_3(x_1) & \varphi_3(x_2) & \varphi_3(x_3) & \varphi_3(x_4)\\ \varphi_4(x_1) & \varphi_4(x_2) & \varphi_4(x_3) & \varphi_4(x_4)\end{vmatrix}=(\vec{a}\times\vec{b}\times\vec{c})\cdot\vec{d}=1. \quad (5)$$

The inverse function $\tilde{\varphi}_i(x_i)$

$$\tilde{\varphi}_m(x_j)=\frac{\Delta_{jm}}{\Delta},$$

which satisfying

$$\sum_{i=1}^{3}\varphi_m(x_i)\tilde{\varphi}_n(x_i)=\sum_{i=1}^{3}\varphi_m(x_i)\frac{\Delta_{in}}{\Delta}=\delta_{mn}. \quad (6)$$

We can use the eigen-solution $a_i(x_i)$ to construct a solution matrix[9]

$$\chi(\tau)=\begin{pmatrix} S_1e^{jx_1\tau} & S_1e^{-jx_1\tau} & T_1e^{jx_3\tau} & -T_1e^{-jx_3\tau} \\ S_2e^{jx_1\tau} & -S_2e^{-jx_1\tau} & T_2e^{jx_3\tau} & T_2e^{-jx_3\tau} \\ S_3e^{jx_1\tau} & S_3e^{-jx_1\tau} & T_3e^{jx_3\tau} & -T_3e^{-jx_3\tau} \\ S_4e^{jx_1\tau} & -S_4e^{-jx_1\tau} & T_4e^{jx_3\tau} & T_4e^{-jx_3\tau} \end{pmatrix}, \tag{7}$$

The obtention of solution matrix $\chi(\tau)$ may be clearly stated, where

$$\varphi_1(x_1)=S_1,\ \varphi_2(x_1)=jS_2,\ \varphi_3(x_1)=S_3,\ \varphi_4(x_1)=jS_4,$$

$$\varphi_1(x_2)=jS_1,\ \varphi_2(x_2)=S_2,\ \varphi_3(x_2)=jS_3,\ \varphi_4(x_2)=S_4,$$

$$\varphi_1(x_3)=T_1,\ \varphi_2(x_3)=jT_2,\ \varphi_3(x_3)=T_3,\ \varphi_4(x_3)=T_4,$$

$$\varphi_1(x_4)=jT_1,\ \varphi_2(x_4)=T_2,\ \varphi_3(x_4)=jT_3,\ \varphi_4(x_4)=jT_4, \tag{8}$$

We can write with clear delimitation Eqs.(4) and (5) between the different vectors $\varphi_i$,

$$\sum_{i=1}^{4}\varphi_i^2(x_1)=S_1^2+S_2^2+S_3^2+S_4^2=N_1^2(\xi^2+2+\eta^2)=1,\ N_1=\frac{1}{\sqrt{\xi^2+2+\eta^2}}=\frac{1}{2},$$

$$\sum_{i=1}^{4}\varphi_i^2(x_4)=T_1^2+T_2^2+T_3^2+T_4^2=N_4^2[(-\frac{\eta\varsigma}{\sqrt{2}})^2+(\frac{\xi\varsigma}{\sqrt{2}})^2+\varsigma^2]=1,\ N_4=\frac{1}{\sqrt{2}\varsigma}, \tag{9}$$

When $\tau\to 0$, corresponding to the starting point $z=0$ of the coupling zone, we have

$$\chi(0)\chi^{-1}(0)=\begin{pmatrix}1&0&0&0\\0&1&0&0\\0&0&1&0\\0&0&0&1\end{pmatrix},\ \chi^{-1}(0)=\begin{pmatrix}S_1&S_2&S_3&S_4\\S_1&-S_2&S_3&-S_4\\T_1&T_2&T_3&T_4\\-T_1&T_2&-T_3&T_4\end{pmatrix}, \tag{10}$$

The import $a_1-a_4$ and output $b_1-b_4$ signals is connected by the transfer function $R(\tau)$. Using $\varphi_m(x_i)$ to construct the matrix $\chi(\tau)$, $\chi^{-1}(\tau)$ and the transvers function $R(\tau)$ [9]

$$R(\tau)=\chi(\tau)\chi^{-1}(0)=\begin{pmatrix}\alpha_1&\mu&\delta&\gamma\\\mu&\alpha_2&\nu&\kappa\\\delta&\nu&\beta_1&\lambda\\\gamma&\kappa&\lambda&\beta_2\end{pmatrix},$$

where

$$\alpha_1=S_1^2\,2\cos(x_1\tau)+T_1^2\,2\cos(x_3\tau),\quad \alpha_2=S_2^2\,2\cos(x_1\tau)+T_2^2\,2\cos(x_3\tau),$$

$$\beta_1=S_3^2\,2\cos(x_1\tau)+T_3^2\,2\cos(x_3\tau),\quad \beta_2=S_4^2\,2\cos(x_1\tau)+T_4^2\,2\cos(x_3\tau),$$

$$\mu = jS_1S_2 2\sin(x_1\tau) + jT_1T_2 2\sin(x_3\tau), \quad \delta = S_1S_3 2\cos(x_1\tau) + T_1T_3 2\cos(x_3\tau)$$

$$\gamma = jS_1S_4 2\sin(x_1\tau) + jT_1T_4 2\sin(x_3\tau), \quad \kappa = S_2S_4 2\cos(x_1\tau) + T_2T_4 2\cos(x_3\tau),$$

$$\nu = jS_2S_3 2\sin(x_1\tau) + jT_2T_3 2\sin(x_3\tau), \quad \lambda = jS_3S_4 2\sin(x_1\tau) + jT_3T_4 2\sin(x_3\tau). \quad (11)$$

## 3. EXAMPLE

The coupling coefficient $\eta \neq 0$, corresponds to two interacted coupled systems[10-12].

$$R(\tau) \to \begin{pmatrix} \frac{\xi^2}{2}\cos\frac{\tau}{\sqrt{2}} + \frac{\eta^2}{2} & j\frac{\xi}{\sqrt{2}}\sin\frac{\tau}{\sqrt{2}} & -\frac{\xi\eta}{2} + \frac{\xi\eta}{2}\cos\frac{\tau}{\sqrt{2}} & 0 \\ j\frac{\xi}{\sqrt{2}}\sin\frac{\tau}{\sqrt{2}} & \cos\frac{\tau}{\sqrt{2}} & j\frac{\eta}{\sqrt{2}}\sin\frac{\tau}{\sqrt{2}} & 0 \\ -\frac{\xi\eta}{2} + \frac{\xi\eta}{2}\cos\frac{\tau}{\sqrt{2}} & j\frac{\eta}{\sqrt{2}}\sin\frac{\tau}{\sqrt{2}} & \frac{\eta^2}{2}\cos\frac{\tau}{\sqrt{2}} + \frac{\xi^2}{2} & 0 \\ 0 & 0 & 0 & 1 \end{pmatrix}. \quad (12)$$

The coupler is depicted in Fig.3. $r_4$ is the radius of the MR2, $L_4$ is the circumference of the MR2, we have

$$a_4 = B_4 b_4 = r_4 e^{j\theta_4}, \quad \theta_4 = \omega L_4/c, \quad (13)$$

And for the MR2, the path $a_2 \to b_2 \to a_3 \to b_3 \to a_2$ indicating

$$a_3 = r_3 e^{j\theta_3} b_2 = B_3 b_2, \quad a_2 = r_2 e^{j\theta_2} b_3 = B_2 b_3, \quad a_4 = r_4 e^{j\theta_4} b_4 = B_4 b_4, \quad (14)$$

The coupling functions may be derived as

$$b_1 = \alpha_1 a_1 + \mu a_2 + \delta a_3 + \gamma a_4, \quad b_2 = \mu a_1 + \alpha_2 a_2 + \nu a_3 + \kappa a_4,$$

$$b_3 = \delta a_1 + \nu a_2 + \beta_1 a_3 + \lambda a_4, \quad b_4 = \gamma a_1 + \kappa a_2 + \lambda a_3 + \beta_2 a_4, \quad (15)$$

Using Eqs.(13)-(15), eliminating $a_2, a_3, a_4$ from Eq.(14), and solving $b_2, b_3, b_4$,

$$(1 - \nu B_3) b_2 - \alpha_2 B_2 b_3 - \kappa B_4 b_4 = \mu a_1,$$

$$-\beta_1 B_3 b_2 + (1 - \nu B_2) b_3 - \lambda B_4 b_4 = \delta a_1,$$

$$-\lambda B_3 b_2 - \kappa B_2 b_3 - (1 - \beta_2 B_4) b_4 = \gamma a_1. \quad (16)$$

Substituting the relations into Eq.(18), we obtain

$$b_2 = \frac{\Delta_2}{\Delta}\ ,\ \ b_3 = \frac{\Delta_3}{\Delta}\ ,\ \ b_4 = \frac{\Delta_4}{\Delta}\ ,$$

where

$$\Delta = \begin{vmatrix} 1-\nu B_3 & -\alpha_2 B_2 & -\kappa B_4 \\ -\beta_1 B_3 & 1-\nu B_2 & -\lambda B_4 \\ -\lambda B_3 & -\kappa B_2 b_3 & 1-\beta_2 B_4 \end{vmatrix},\ \ \Delta_2 = \begin{vmatrix} \mu a_1 & -\alpha_2 B_2 & -\kappa B_4 \\ \delta a_1 & 1-\nu B_2 & -\lambda B_4 \\ \gamma a_1 & -\kappa B_3 b_2 & 1-\beta_2 B_4 \end{vmatrix},$$

$$\Delta_3 = \begin{vmatrix} 1-\nu B_3 & \mu a_1 & -\kappa B_4 \\ -\beta_1 B_3 & \delta a_1 & -\lambda B_4 \\ -\lambda B_3 & \gamma a_1 & 1-\beta_2 B_4 \end{vmatrix},\ \ \Delta_4 = \begin{vmatrix} 1-\nu B_3 & -\alpha_2 B_2 & \mu a_1 \\ -\beta_1 B_3 & 1-\nu B_2 & \delta a_1 \\ -\lambda B_3 & -\kappa B_3 b_2 & \gamma a_1 \end{vmatrix}, \qquad (17)$$

The transmission is

$$Tr = \tau = \left|\frac{b_1}{a_1}\right|^2 = \left|\alpha_1 + \mu B_2 b_3 + \delta B_3 b_2 + \gamma B_4 b_4\right|^2 = \left|\alpha_1 + \mu B_2 \frac{\Delta_2}{a_1\Delta} + \delta B_3 \frac{\Delta_3}{a_1\Delta} + \gamma B_4 \frac{\Delta_4}{a_1\Delta}\right|^2 .(18)$$

If the coupling between the waveguides $a_2 b_2$ and $a_3 b_3$ is so weak, the coupling coefficient $\eta \to 0$ , corresponds to two independent coupled systems. As shown in Eq.(8), $\delta \to 0, \kappa \to 0, \gamma \to 0, \nu \to 0$ . The general transfer matrix $R(\tau)$ solutions

$$R(\tau) = \chi(\tau)\chi^{-1}(0) = \begin{pmatrix} \alpha_1 & \mu & \delta & \gamma \\ \mu & \alpha_2 & \nu & \kappa \\ \delta & \nu & \beta_1 & \lambda \\ \gamma & \kappa & \lambda & \beta_2 \end{pmatrix} \to \begin{pmatrix} \alpha_1 & \mu & 0 & 0 \\ \mu & \alpha_2 & 0 & 0 \\ 0 & 0 & \beta_1 & \lambda \\ 0 & 0 & \lambda & \beta_2 \end{pmatrix}, \qquad (19)$$

In terms of Eq.(12), and the coupled mode equations, the transmission of the two MRs are $t_1, t_2$ , the loss of the two MRs are $k_1$ , $k_2$ , respectively.

$$t_2 = \alpha_1 = \cos(x_1\tau)\,,\quad jk_2 = \mu = j\sin(x_1\tau)\,,$$

$$t_1 = \beta_1 = \beta_2 = \cos(x_3\tau)\,,\quad jk_1 = \lambda = j\sin(x_3\tau)\,,$$

$$k_1^2 + t_1^2 = 1,\qquad k_2^2 + t_2^2 = 1. \qquad (20)$$

Eq.(15) becomes to

$$b_3 = \frac{\Delta_3}{\Delta} \to \frac{jk_2 a_1 B_3 (t_1 - B_4)}{1 - t_1 B_4 - t_2 (t_1 - B_4) B_2 B_3} = \frac{jk_2 a_1 \tau_1 B_3}{1 - t_2 B_2 \tau_1 B_3}\,,\quad \tau_1 = \frac{t_1 - B_4}{1 - t_1 B_4}\,,$$

where

$$\Delta_2 \to \begin{vmatrix} \mu a_1 & -\alpha_2 B_2 & 0 \\ 0 & 1 & -\lambda B_4 \\ 0 & 0 & 1-\beta_2 B_4 \end{vmatrix} = \mu a_1 (1-\beta_2 B_4) = \mu a_1 (1-t_1 B_4) \ ,$$

$$\Delta_3 \to \begin{vmatrix} 1 & \mu a_1 & 0 \\ -\beta_1 B_3 & 0 & -\lambda B_4 \\ -\lambda B_3 & 0 & 1-\beta_2 B_4 \end{vmatrix} = \mu a_1 (\beta_1 B_3 (1-\beta_2 B_4) + \lambda^2 B_3 B_4)$$

$$= jk_2 a_1 B_3 (t_1 (1-t_1 B_4) - (1-t_1^2) B_4) = jk_2 a_1 B_3 (t_1 - B_4) ,$$

$$\Delta \to \begin{vmatrix} 1 & -\alpha_2 B_2 & 0 \\ -\beta_1 B_3 & 1 & -\lambda B_4 \\ -\lambda B_3 & 0 & 1-\beta_2 B_4 \end{vmatrix} = 1-\beta_2 B_4 - \alpha_2 \lambda^2 B_2 B_3 B_4 - \alpha_2 \beta_1 B_2 B_3 (1-\beta_2 B_4)$$

$$= 1 - t_1 B_4 + t_2 (1-t_1^2) B_2 B_3 B_4 - t_2 t_1 B_2 B_3 (1-t_1 B_4) = 1 - t_1 B_4 - t_2 t_1 B_2 B_3 + t_2 B_4 B_2 B_3 . \qquad (21)$$

The transmission is

$$Tr = \tau = \left| \frac{b_1}{a_1} \right|^2 \to \left| \frac{\alpha_1 a_1 + \mu a_2}{a_1} \right|^2 = \left| t_2 + \frac{jk_2 B_2 b_3}{a_1} \right|^2 = \left| t_2 - \frac{k_2^2 \tau_1 B_2 B_3}{1 - t_2 \tau_1 B_2 B_3} \right|^2 = \tau_2 . \qquad (22)$$

The results of $\tau_1$ , $\tau_2$ can be obtained from Eq.(22). Tthe parameter $B_4$ corresponds to $a_1 e^{j\theta_1}$ , and $B_2 B_3$ corresponds to $a_2 e^{j\theta_2}$ in Smith's formula[5].

## 4. NUMERICAL SIMULATIONS

In this section, we evaluate numerically the influence of interaction between MRs on the transmission ( $Tr$ ) and the phase ( $ph = Arg(\tau)$ ) of the coupler in Fig.1, where $t_1$ is the transmission coefficient between the bus waveguide and MRR1, $t_2$ is the transmission coefficient between the MRR1 and MRR2, $\theta_1 = \theta_2 = \theta$ is the phase shift after the light circulate along the MR, $a_1$ is the absorption coefficient between the bus waveguide and MR1, $a_2$ is the absorption coefficient between the MR1 and MR2.

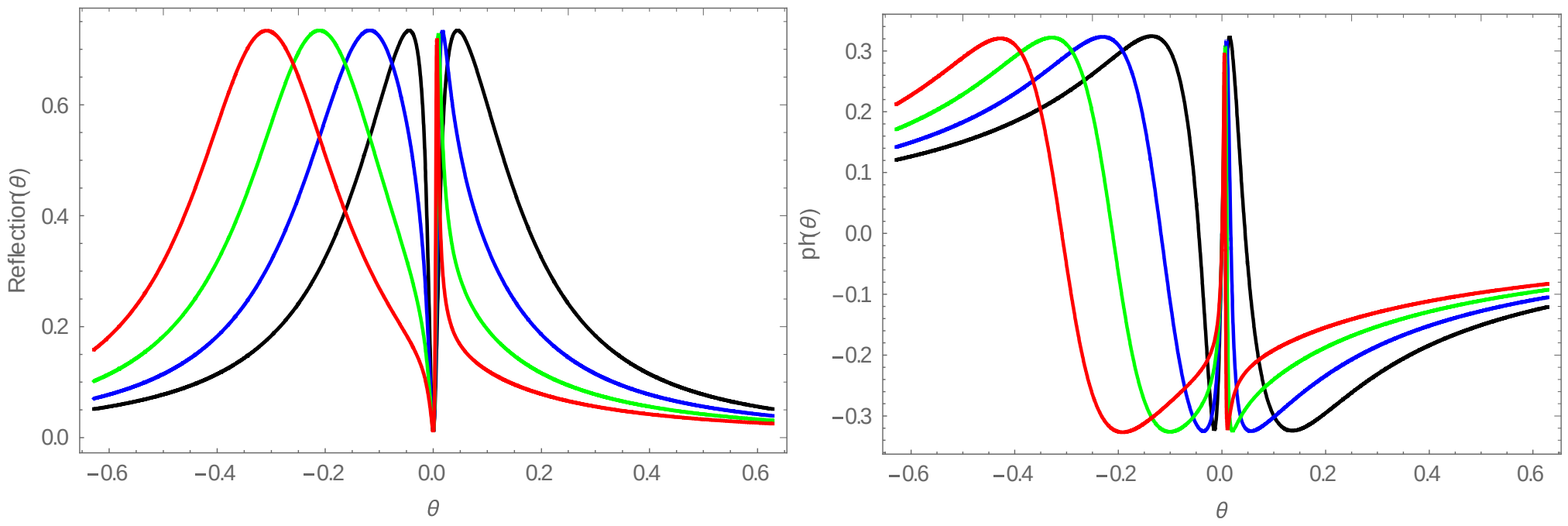

(a) The transmission spectra (b) the phase shift spectra

**Fig. 2.** Numerical solution for an asymmetric coupler with the asymmetric factors $\eta = 0$ (black), $\eta = 0.1$ (blue), $\eta = 0.2$ (green), $\eta = 0.3$ (red) (a) the reflection spectra $1 - Tr$ and (b) phase spectra $ph$ evolves with $\theta \in (-0.2\pi, 0.2\pi) rad$ predicted by parameters from Ref.[3] for the transmission coefficients $t_1 = 0.999$, $t_2 = 0.96$ and the absorption coefficients $a_1 = 0.9999$, $a_2 = 0.88$.

As shown in Fig.4(a), when asymmetric factor is zero, the transmission spectrum is EIT-like shape. Continuing to increasing of the asymmetric factors from 0 to 0.3, the spectrum has an increasingly asymmetric, and the position of the first resonance peak occurs left shift, the bandwidth are broaden[5]. while the bandwidth of the second resonance peak is disappear, the transmission spectrum turns to Fano shape. As shown in Fig.4(b), the trend of change of the phase spectrum is exactly the same as the transmission spectrum, half in positive region, half in negative region.

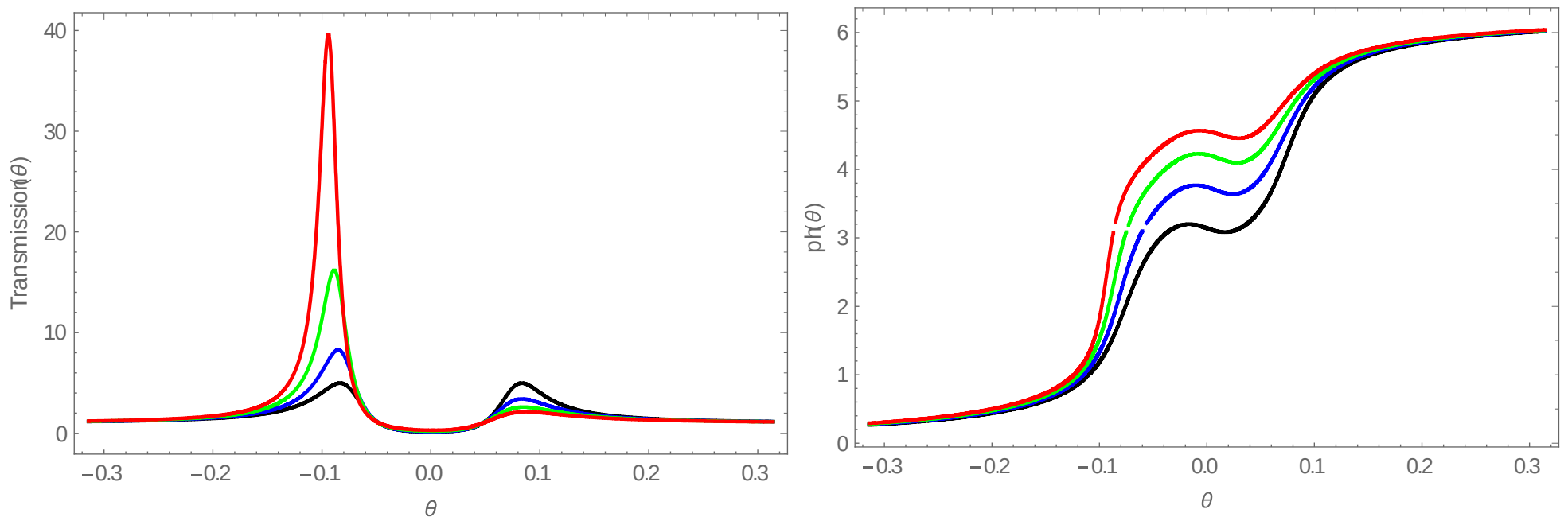

(a) The transmission spectra (b) the phase shift spectra

**Fig. 3.** Numerical solution for an asymmetric coupler with the coupling factors $\eta = 0$ (black), $\eta = 0.01$ (blue), $\eta = 0.02$ (green), $\eta = 0.03$ (red) (a) transmission spectra $Tr$ and (b) phase shift spectra $ph$ evolves with $\theta \in (-0.1\pi, 0.1\pi) rad$ predicted by parameters from Ref.[4] for the transmission coefficient $t_1 = 0.994$ , $t_2 = 0.96$ and the absorption coefficients $a_1 = 0.9$, $a_2 = 1.1$.

As shown in Fig.6(a), we continue to increasing of the coupling factors from symmetric to asymmetric, we use smaller coupling factors $\eta$ and a larger $a_2$, besides the resonant wavelength position of the transmission spectrum unchanged, the height of the left peak increases greatly and the height of the right peak nearly disappears, the shape of the EIT-like spectrum will disappear. As shown in Fig.6(b), The phase spectrum is all in positive region.

## 5. CONCLUSIONS

In this paper, we use MRRs and bus wave-guides to establish an asymmetric coupler. The principle is illustrated by transfer matrix method and coupled mode theory, EIT-like spectrum and Fano spectrum generated by bound states in the continuum (BIC) excitation. The spectrum can be independently tuned by changing micro-ring parameters as shown in Figs.2-3. Our system has many potential applications in the field of optical switches, fast or slow light, optical communication.

## APPENDIX A: CALCULATION OF THE DETERMINATES

Considering the mode coupling between non-adjacent waveguides and loss, the coupled-mode equation of a general $3\times3$ asymmetric coupler can be described by

$$\begin{pmatrix} \frac{\partial a_1}{\partial z} \\ \frac{\partial a_2}{\partial z} \\ \frac{\partial a_3}{\partial z} \end{pmatrix} = \begin{pmatrix} 0 & j\frac{K}{2}\xi & j\frac{K}{2}\varsigma \\ j\frac{K}{2}\xi & 0 & j\frac{K}{2}\eta \\ j\frac{K}{2}\varsigma & j\frac{K}{2}\eta & 0 \end{pmatrix} \begin{pmatrix} a_1 \\ a_2 \\ a_3 \end{pmatrix}, \tag{A.1}$$

The second micro-ring $a_3 \to b_3$, $a_3 = c_3 e^{j\theta_3} b_3 = B_3 b_3$, light beam propagating in counterclockwise direction, we have the transmission is

$$T = \left|\frac{b_1}{a_1}\right|^2 = \left|\frac{\Delta_1}{a_1\Delta}\right|^2 = \left|\frac{\alpha_1 + D_{33}B_2 + D_{22}B_3 + DB_2B_3}{1-\beta_2 B_2 - \gamma_3 B_3 + D_{11}B_2B_3}\right|^2,$$

where

$$D=\begin{vmatrix}\alpha_1 & \alpha_2 & \alpha_3\\ \beta_1 & \beta_2 & \beta_3\\ \gamma_1 & \gamma_2 & \gamma_3\end{vmatrix},\ D_{11}=\begin{vmatrix}\beta_2 & \beta_3\\ \gamma_2 & \gamma_3\end{vmatrix},\ D_{22}=\begin{vmatrix}\alpha_1 & \alpha_3\\ \gamma_1 & \gamma_3\end{vmatrix},\ D_{33}=\begin{vmatrix}\alpha_1 & \alpha_2\\ \beta_1 & \beta_2\end{vmatrix}, \qquad \text{(A.2)}$$

The determinates $D$ , $R_{11}$, $R_{22}$, $R_{33}$ defined in Eq.(A.2) belongs the determinates of transfer matrix $R$, $R_{11}$, $R_{22}$, $R_{33}$. Let $L=e^{ix_1\tau}$, $L=e^{ix_2\tau}$, $L=e^{ix_3\tau}$ write Eq. (8) in the form

$$\begin{aligned}
\alpha_1 &= \varphi_1(x_1)\tilde{\varphi}_1(x_1)e^{jx_1\tau}+\varphi_1(x_2)\tilde{\varphi}_1(x_2)e^{jx_2\tau}+\varphi_1(x_3)\tilde{\varphi}_1(x_3)e^{jx_3\tau}=a_{11}^{l}L+a_{11}^{m}M+a_{11}^{n}N,\\
\alpha_2 &= \varphi_2(x_1)\tilde{\varphi}_1(x_1)e^{jx_1\tau}+\varphi_2(x_2)\tilde{\varphi}_1(x_2)e^{jx_2\tau}+\varphi_2(x_3)\tilde{\varphi}_1(x_3)e^{jx_3\tau}=a_{21}^{l}L+a_{21}^{m}M+a_{21}^{n}N,\\
\alpha_3 &= \varphi_3(x_1)\tilde{\varphi}_1(x_1)e^{jx_1\tau}+\varphi_3(x_2)\tilde{\varphi}_1(x_2)e^{jx_2\tau}+\varphi_3(x_3)\tilde{\varphi}_1(x_3)e^{jx_3\tau}=a_{31}^{l}L+a_{31}^{m}M+a_{31}^{n}N,\\
\beta_1 &= \varphi_1(x_1)\tilde{\varphi}_2(x_1)e^{jx_1\tau}+\varphi_1(x_2)\tilde{\varphi}_2(x_2)e^{jx_2\tau}+\varphi_1(x_3)\tilde{\varphi}_2(x_3)e^{jx_3\tau}=a_{12}^{l}L+a_{12}^{m}M+a_{12}^{n}N,\\
\beta_2 &= \varphi_2(x_1)\tilde{\varphi}_2(x_1)e^{jx_1\tau}+\varphi_2(x_2)\tilde{\varphi}_2(x_2)e^{jx_2\tau}+\varphi_2(x_3)\tilde{\varphi}_2(x_3)e^{jx_3\tau}=a_{22}^{l}L+a_{22}^{m}M+a_{22}^{n}N, \quad \text{(A.3)}\\
\beta_3 &= \varphi_3(x_1)\tilde{\varphi}_2(x_1)e^{jx_1\tau}+\varphi_3(x_2)\tilde{\varphi}_2(x_2)e^{jx_2\tau}+\varphi_3(x_3)\tilde{\varphi}_2(x_3)e^{jx_3\tau}=a_{32}^{l}L+a_{32}^{m}M+a_{32}^{n}N,\\
\gamma_1 &= \varphi_1(x_1)\tilde{\varphi}_3(x_1)e^{jx_1\tau}+\varphi_1(x_2)\tilde{\varphi}_3(x_2)e^{jx_2\tau}+\varphi_1(x_3)\tilde{\varphi}_3(x_3)e^{jx_3\tau}=a_{13}^{l}L+a_{13}^{m}M+a_{13}^{n}N,\\
\gamma_2 &= \varphi_2(x_1)\tilde{\varphi}_3(x_1)e^{jx_1\tau}+\varphi_2(x_2)\tilde{\varphi}_3(x_2)e^{jx_2\tau}+\varphi_2(x_3)\tilde{\varphi}_3(x_3)e^{jx_3\tau}=a_{23}^{l}L+a_{23}^{m}M+a_{23}^{n}N,\\
\gamma_3 &= \varphi_3(x_1)\tilde{\varphi}_3(x_1)e^{jx_1\tau}+\varphi_3(x_2)\tilde{\varphi}_3(x_2)e^{jx_2\tau}+\varphi_3(x_3)\tilde{\varphi}_3(x_3)e^{jx_3\tau}=a_{23}^{l}L+a_{23}^{m}M+a_{23}^{n}N.
\end{aligned}$$

$$D_{11}=\begin{vmatrix}\beta_2 & \beta_3\\ \gamma_2 & \gamma_3\end{vmatrix}=\beta_2\gamma_3-\beta_3\gamma_2$$

$$=(a_{22}^{l}L+a_{22}^{m}M+a_{22}^{n}N)(a_{33}^{l}L+a_{33}^{m}M+a_{33}^{n}N)-(a_{23}^{l}L+a_{23}^{m}M+a_{23}^{n}N)(a_{32}^{l}L+a_{32}^{m}M+a_{32}^{n}N)$$

$$=(a_{22}^{l}a_{33}^{l}-a_{23}^{l}a_{32}^{l})L^2+(a_{22}^{m}a_{33}^{m}-a_{23}^{m}a_{32}^{m})M^2+(a_{22}^{n}a_{33}^{n}-a_{23}^{n}a_{32}^{n})N^2$$
$$+(a_{22}^{l}a_{33}^{m}+a_{22}^{m}a_{33}^{l}-a_{23}^{l}a_{32}^{m}-a_{23}^{m}a_{32}^{l})LM+(a_{22}^{l}a_{33}^{n}+a_{22}^{n}a_{33}^{l}-a_{23}^{l}a_{32}^{n}-a_{23}^{n}a_{32}^{l})LN$$
$$+(a_{22}^{m}a_{33}^{n}+a_{22}^{n}a_{33}^{m}-a_{23}^{m}a_{32}^{n}-a_{23}^{n}a_{32}^{m})MN \qquad , \qquad \text{(A.4)}$$

Owing to $a_{22}^{l}a_{33}^{l}-a_{23}^{l}a_{32}^{l}=\varphi_2(x_1)\tilde{\varphi}_2(x_1)\varphi_3(x_1)\tilde{\varphi}_3(x_1)-\varphi_2(x_1)\tilde{\varphi}_3(x_1)\varphi_3(x_1)\tilde{\varphi}_2(x_1)=0$, the first term vanishes. For the same reason, the second and third terms all vanish.

The coefficient of $LM$ in Eq. (A.2) may be written as

$$\begin{aligned}
&a_{22}^{l}a_{33}^{m}+a_{22}^{m}a_{33}^{l}-a_{23}^{l}a_{32}^{m}-a_{23}^{m}a_{32}^{l}\\
&=\varphi_2(x_1)\tilde{\varphi}_2(x_1)\varphi_3(x_2)\tilde{\varphi}_3(x_2)+\varphi_2(x_2)\tilde{\varphi}_2(x_2)\varphi_3(x_1)\tilde{\varphi}_3(x_1)\\
&\quad -\varphi_2(x_1)\tilde{\varphi}_3(x_1)\varphi_3(x_2)\tilde{\varphi}_2(x_2)-\varphi_2(x_2)\tilde{\varphi}_3(x_2)\varphi_3(x_1)\tilde{\varphi}_2(x_1)\\
&=(\varphi_2(x_1)\varphi_3(x_2)-\varphi_2(x_2)\varphi_3(x_1))(\tilde{\varphi}_2(x_1)\tilde{\varphi}_3(x_2)-\tilde{\varphi}_2(x_2)\tilde{\varphi}_3(x_1))\\
&=\tilde{\varphi}_1(x_3)\varphi_1(x_3)=\tilde{\varphi}_1(x_3)\varphi_1(x_3)=\tilde{\varphi}_1(x_3)\varphi_1(x_3) \qquad , \qquad \text{(A.5)}
\end{aligned}$$

Finally, we derive the equation for $D_{11}$ [14]

$$\begin{aligned} D_{11} &= \varphi_1(x_1)\tilde{\varphi}_1(x_1)MN + \varphi_1(x_2)\tilde{\varphi}_1(x_2)LN + \varphi_1(x_3)\tilde{\varphi}_1(x_3)LM \\ &= \varphi_1(x_1)\tilde{\varphi}_1(x_1)L^{-1} + \varphi_1(x_2)\tilde{\varphi}_1(x_2)M^{-1} + \varphi_1(x_3)\tilde{\varphi}_1(x_3)N^{-1} = \alpha_1^*, \end{aligned} \tag{A.6}$$

Similarly, the derivations for $D_{22}, D_{33}$ are [14]

$$\begin{aligned} D_{22} &= \varphi_2(x_1)\tilde{\varphi}_2(x_1)L^{-1} + \varphi_2(x_2)\tilde{\varphi}_2(x_2)M^{-1} + \varphi_2(x_3)\tilde{\varphi}_2(x_3)N^{-1} = \beta_2^*, \\ D_{33} &= \varphi_3(x_1)\tilde{\varphi}_3(x_1)L^{-1} + \varphi_3(x_2)\tilde{\varphi}_3(x_2)M^{-1} + \varphi_3(x_3)\tilde{\varphi}_3(x_3)N^{-1} = \gamma_3^*. \end{aligned} \tag{A.7}$$

Using the expansion to calculate the determinate $D$, the coefficients of terms $L^3, L^2M, \ldots,$ can be proved similarly being vanishing as the coefficients of terms $L^2, M^2, N^2$ in Eq. (A.2). The only non-vanishing coefficient is the term $LMN = e^{i(x_1+x_2+x_3)\tau} = 1$ due to $x_1 + x_2 + x_3 = 0$. Thus, we can calculate $D$ by putting $\tau = 0$ [14].

$$D = \begin{vmatrix} \alpha_1 & \alpha_2 & \alpha_3 \\ \beta_1 & \beta_2 & \beta_3 \\ \gamma_1 & \gamma_2 & \gamma_3 \end{vmatrix} = \left|\chi(\tau)\chi^{-1}(0)\right| = \left|\chi(0)\chi^{-1}(0)\right| = 1. \tag{A.8}$$

**Funding**

Financial support from the project funded by the State Key Laboratory of Quantum Optics and Quantum Optics Devices, Shanxi University, Shanxi, China (Grants No.KF202004 and No. KF202205).

**Disclosures.** The authors declare no conflicts of interest.

**Data Availability.** Data underlying the results presented in this paper are not publicly available at this time but may be obtained from the authors upon reasonable request.